# Prediction of the Atmospheric Fundamental Parameters from Stellar Spectra Using Artificial Neural Network

Y. A. Azzam, M. I. Nouh, and A. A. Shaker

Department of Astronomy, National Research Institute of Astronomy and Geophysics (NRIAG), 11421 Helwan, Cairo, Egypt
e-mail: y.azzam@nriag.sci.eg

**Abstract:**

Innovation in the ground and space-based instruments has taken us into a new age of spectroscopy, in which a large amount of stellar content is becoming available. So, automatic classification of stellar spectra became subjective in recent years due to the availability of large observed spectral database as well as the theoretical spectra. In the present paper, we develop an Artificial Neural Network (ANN) algorithm for automated classification of stellar spectra. The algorithm has been applied to extract the fundamental parameters of some hot helium rich white dwarf stars observed by the Sloan Digital Sky Survey (SDSS) and B-type spectra observed at Onderjove observatory. We compared the present fundamental parameters and those from a minimum distance method to clarify the accuracy of the present algorithm where we found that, the predicted atmospheric parameters for the two samples are in good agreement for about 50% of the samples. A possible explanation for the discrepancies found for the rest of the samples is discussed.

## 1. Introduction

A huge quantity of stellar spectra is found in large scale sky surveys. This large number of stellar spectra ensures that spectral data must be parameterized automatically, which allows investigating carefully the characteristics of atmospheric parameters.

In the last decade of the previous century, machine learning techniques have been used to automate stellar spectra. One of these techniques, Artificial Neural networks (ANNs) have acquired a very good reputation in this operation. It is also well known that ANNs have acquired an eminent role in many areas of human activity over the past decades and have found applications in a wide range of scientific issues, including astronomy, microbiology, geophysics and the environment sciences, Ozard and Morbey (1993), Almeida and Noble (2000), Tagliaferri et. Al. (2003), Faris et al. (2014), Elminir et al. (2007). It was commonly used in the areas of pattern recognition, data classification, prediction, function approximation, signal processing, medical diagnosis, modeling, and control, etc., El-Mallawany et al. (2014), Al-Shayea (2011), Leshno et al. (1993), Lippmann (1989), Zhang (2000). The ANN is mathematical models hinted by biological neural systems and composed of neuron models that are connected in a parallel and distributed style to imitate the knowledge acquisition and information processing of the human nervous system. The computations of the ANNs are performed at a very high speed because of their massively parallel nature. They possess the capabilities of learning and self-organization that can memorize and pick up a mapping between an input and an output vector space and synthesize an associative memory which recovers the correct output when the input is introduced and generalizes when new inputs are introduced, Basheer and Hajmeer (2000). ANN was used in astronomy in many fields such as adaptive optics, star/galaxy separation as well as galaxy classification.

Because of their excellent properties of fault tolerance, self-learning, adaptivity, nonlinearity, ANNs were used extensively for stellar spectral classifications, Weaver and Torres-Dodgen (1995), Weaver and Torres-Dodgen (1997), Torres-Dodgen (2000), Snider (2001), Gulati and Gupta (1995), Vieira and Ponz (1998), Bailer-Jones et al. (1997). In these researches, different spectral classifications for different wavelength ranges and different spectral resolutions were performed for different spectral types by the use of ANN techniques. The goal was to use ANN instead of human experts to automatically classify stellar spectra in large spectral surveys, which were motivated by the advent of a combination of digital computers, high efficiency CCD cameras, and spectrographs with fiber optics multiplexing. At the time of those researches, the processing power of computers was limited. As a result, various researchers used Principal Components Analysis (PCA) with ANN to reduce and compress the amount of data feed to the network, Serra et al. (1993), Bailer-Jones et al. (1996), Singh et al. (1998), Bailer-Jones et al. (1998), Tagliaferri et al. (1999). Nowadays, the processing power of computers has doubled many times and the advancement occurred in the resolving power of telescopes as well as attached instruments, together with the advancement in detectors' efficiencies have the motivation of producing large scale sky surveys. In these surveys, an enormous number of stellar spectra are found. Many stellar spectra mean that spectral data are properly parameterized so that the stellar fundamental parameters can be thoroughly examined. Examples of such surveys are the Large Sky Area MultiObject Fiber Spectroscopic Tele-secure (LAMOST; Zhao et al. 2006; Luo et al. 2015; Cui et al. 2012), and Gaia-ESO Survey (Gilmore et al. 2012; Randichet et al. 2013), Sloan Digital Sky Survey (SDSS, York et al. 2000; Alam et al. 2015; Ahn et al. 2012) and STELB stellar library (Borgne et al., 2003).

A possible approach for spectral classification using automatic methods is to tie the observed spectrum up to the synthetic spectrum based on theoretical stellar atmospheric models to understand the physical phenomena in stars. Using theoretical stellar spectra, Kurucz' (1992) for example, instead of empirical libraries has the advantage that they can be calculated for a dense grid of fundamental parameters (metallicity, gravity, effective temperatures), thus avoiding interpolation errors and calibrations. In this sense, for example, Gulati et al. (1997) used ANN to determine the effective temperatures for G-K dwarfs and compared them with those given in Gray and Corbally (1994). Li et al. (2017) used deep learning techniques to estimate the atmospheric parameters from stellar spectra.

In the present paper, we introduce an ANN approach to predict the fundamental stellar parameters. For the training stage, we use two grids of synthetic spectra, the first for the DO white dwarfs and the second for the B-type stars. The algorithm is tested by deriving fundamental parameters for the list of observed DO white dwarfs retrieved from SDSS release 4. The spectra of the B-type stars are retrieved from the archive of the Onderjove observatory. The rest of the paper is organized as follows: Section 2 deals with the principle of neural network algorithm, whereas section 3 is devoted to the minimum distance method algorithm. In section 4, an explanation of synthetic spectra and data preprocessing for both DO white dwarf and B-type stars are explored. Section 5 deals with the results and discussion of the application of ANN for stellar parameterization, and the conclusion is given in section 6.

## 2. Neural Network Algorithm

In ANN, the main processing unit which can carry out localized information and can process a local memory is the neuron. The net input ($x$) at each neuron is calculated by adding the weights it receives to get a weighted sum of those inputs and add it with a bias ($b$). The incorporation of the bias in the process is to permit offsetting the activation function from zero,

$$x = (w_{1,1}.p_1 + w_{1,2}.p_2 + \rightleftharpoons ... \rightleftharpoons + w_{1,j}.p_j) + b \qquad (1)$$

Then the net input $(x)$ is passed through an activation function, resulting in the output of the neuron $(y_j)$. The activation function used to transform input to an output level in the range of 0.0 to 1.0 is a nonlinear sigmoid function which is a conventional sigmoid function represented by the following expression:

$$y_j = \frac{1}{1+e^{-x}} \tag{2}$$

A comparison of the output $y_j$ at the output layer with the target output $t_j$ is implemented using an error function that has the following form:

$$\delta_j = y_j(t_j - y_j)(1 - y_j) \tag{3}$$

For the hidden layer, the error function takes the form:

$$\delta_j = y_j(1 - y_j)\sum_k \delta_k w_k \tag{4}$$

where $\delta_j$ is the output layer error term, and $w_k$ is the weight between the hidden and output layers. To update the weight of each connection, the error is replicated backward from the output layer to the input layer as follows:

$$w_{ji}(t+1) = w_{ji}(t) + \eta \delta_j y_j + \beta(w_{ji}(t) - w_{ji}(t-1)) \tag{5}$$

Learning rate $\eta$ has to be chosen such that it is neither very small leading to a slow rate of convergence nor too large leading to overshooting. The constant β is called the momentum factor and is used to speed up the convergence of the back-propagation learning algorithm error. This term has the effect of adding a fraction of the most recent weight adjustment to the current weight adjustments. Both η and β terms are assigned at the beginning of the training phase and decide the network stability and speed, Elminir (2007), Basheer and Hajmeer (2000). For each input pattern, the process is repeated until the network output error is reduced to a pre-assigned threshold value. The final weights are frozen and used to obtain fundamental stellar parameters $T_{eff}$, $\log g$ during the test session. To assess the success and quality of the training, an error is calculated for the whole batch of training patterns. In this paper, Root Mean Squared error ($E_{rms}$) is used which is defined as:

$$E_{rms} = \frac{1}{n}\sqrt{\sum_{n=1}^{n}(t_j - y_j)^2}, \tag{6}$$

where n is the number of training patterns. An error of zero would indicate that all the output patterns calculated by the ANN match the expected values perfectly and that the ANN is well trained. We used the feed forward neural network, as shown in Fig 1, to simulate the fundamental stellar parameters $T_{eff}$, $\log g$. It has a hierarchical structure that consists of an input layer, hidden layer and output layer with only interconnections between the neurons in subsequent layers, and signals can propagate only from the input layer to the output layer through the hidden layer.

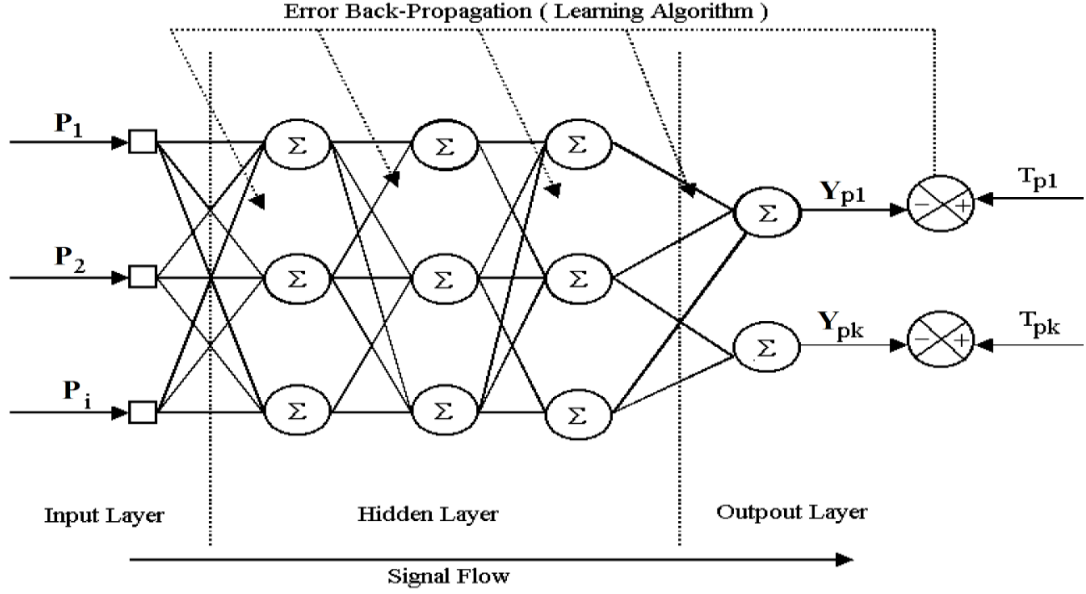

Figure 1. ANN architecture proposed to simulate the fundamental stellar parameters.

### 3. Minimum Distance Method

The minimum distance method (MDM) is widely used to determine the fundamental stellar parameters by comparing the observed spectra with the grids of theoretical spectra. The Euclidian distance between observed $O_i$ and template $T_i$ fluxes could be written as (Allende Prieto. 2004)

$$d = \sum p_i u_i \tag{7}$$

Where

$$u_i = \left[O_i - T_i(x)\right]^2 \tag{8}$$

Where $x$ is the vector of the fundamental parameters i.e. $T_{eff}$, $\log g$.

The weight $p_i$ could be given by

$$p_i = \sum_i \frac{1}{I(x_i)} \left|\frac{\partial u_i}{\partial x_i}\right| \tag{9}$$

Where

$$I(x_i) = \sum_i \left|\frac{\partial u_i}{\partial x_i}\right| \tag{10}$$

In the present calculations, we take $p_i = 1$, so we didn't need the interpolation between the flux grids.

## 4. Synthetic Spectra and Data Preprocessing

### 4.1 Synthetic Spectra

We used the DO white dwarf atmospheric grid computed by Nouh and Fouda (2007). We modeled the structure of the atmospheric using the TLUSTY code (version 200; Hubeny 1988; Hubeny and Lanz 1992, 1995, 2003; Lanz and Hubeny 2001; Lanz et al. 2003). The atmosphere is assumed as plane-parallel, in radiative and hydrostatic, and the convection is treated with the mixing length theory. Departures from local thermodynamic equilibrium (LTE) are allowed for an arbitrary set of atomic and ionic energy levels, Lanz and Hubeny (2006). The effective temperatures span the range 40000-120000 $K^0$ with step 2500 $K^0$ and the surface gravities span the range log g = 7 - 8.5 with step 0.25.

The general spectrum synthesis code SYNSPEC (version 48, Hubeny and Lanz, 2003) was used to synthesis the spectra in the wavelength range $\lambda\lambda$3000-7000 $A^0$ with a sampling 0.1 $A^0$. The input model atmospheres to SYNSPEC is taken from TLUSTY. The radiative transfer equation is solved wavelength by wavelength in a specified wavelength range and with a specified wavelength resolution. To bring the synthetic spectra to the resolution of the observed SDSS spectra, we used a Gaussian profile with FWHM=3 $A^0$. The calculations were performed for a nearly pure helium atmosphere with He/H=1000 by numbers, and the heavier elements have been neglected. Figure 2 shows the normalized flux shifted by an arbitrary value for more clarity.

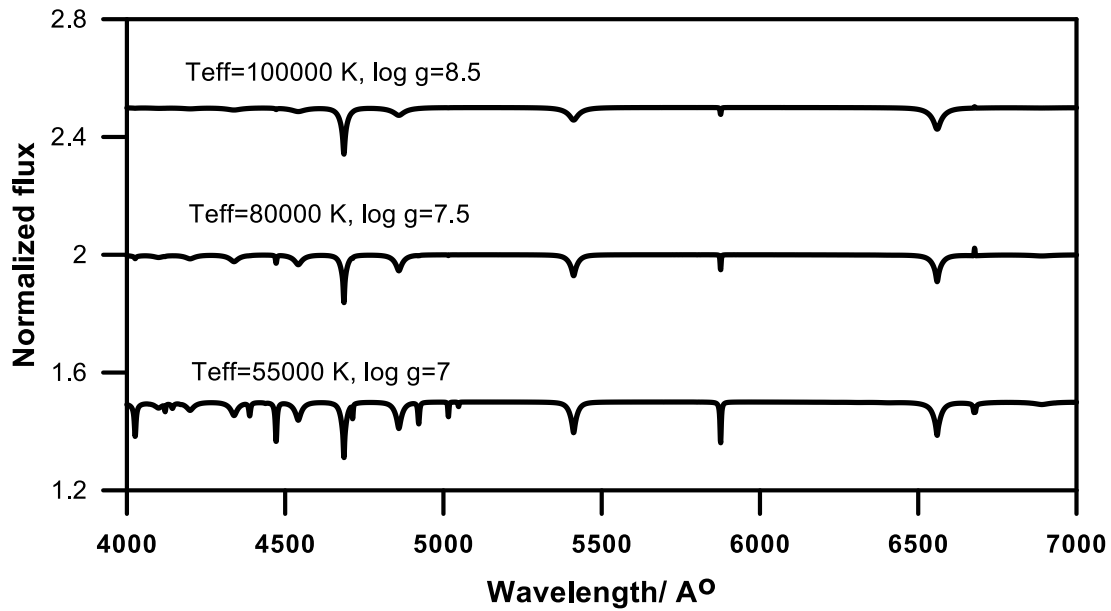

**Figure 2: Normalized synthetic spectra of the DO white dwarfs used for training the ANN. Spectra are labeled with effective temperature and surface gravity.**

The atmospheric models of the B-type stars are adopted from Lanz and Hubeny (2005) using TLUSTY code version 200 (Hubeny 1988& Hubeny and Lanz 1992& 1995& 2003; Lanz and Hubeny 2001; Lanz et al. 2003). We adopt the grid with solar metallicity. The models span the range in the effective temperatures Teff= 15000-30000 K with step 1000 K and the surface gravity span the range log g=1.75-4.75 with 0.25 dex step. Also, we used the code SYNSPEC to synthesis the spectra. The input model atmosphere is taken from TLUSTY models. The resulting spectra are reduced to the resolution of the observed Onderjov spectra using a Gaussian profile with FWHM=0.25 $A^0$. Figure 3 shows the normalized flux shifted by an arbitrary value for more clarity.

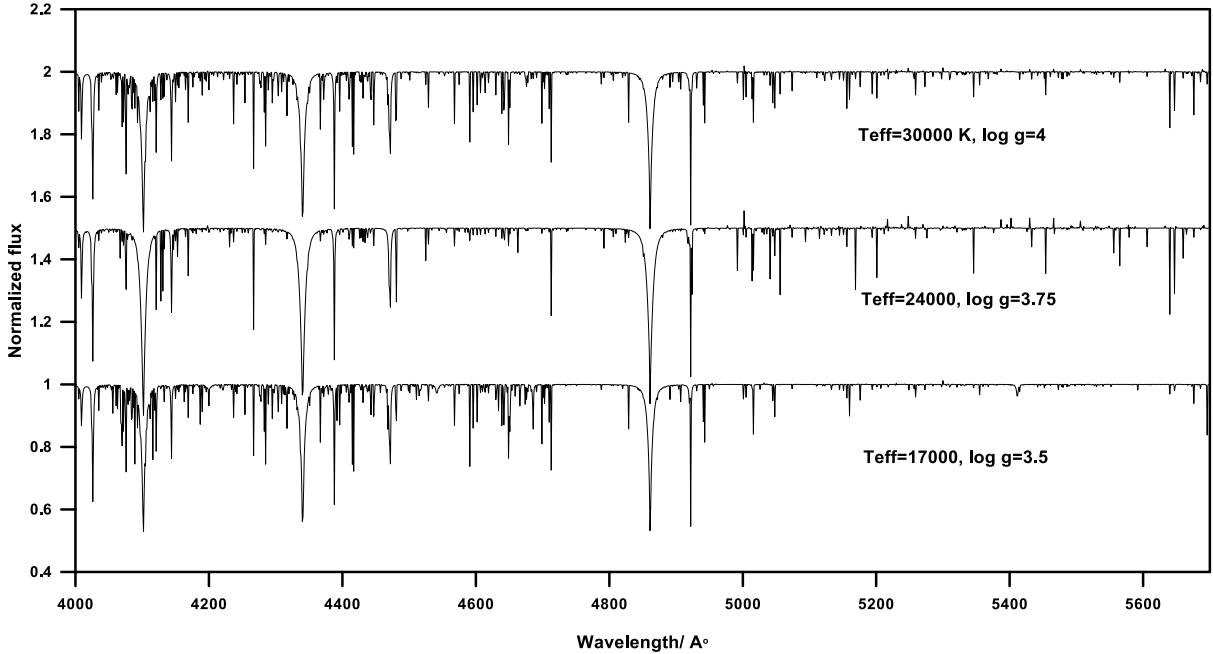

**Figure 3: Normalized synthetic spectra of the B-type stars used to train the ANN. Spectra are labeled with effective temperature and surface gravity**.

### 4.2  Data Preprocessing and Unification

The use of ANN algorithm to automate the stellar spectral classification process necessitates the use of uniform datasets. This requires the unification of the whole training dataset being used to train the neural network as well as the test dataset used to verify the quality of the proposed algorithm. More specifically, the training and test datasets used must be unified to the same stellar wavelength range with identical starting and ending values, same spectral resolutions, and their fluxes must be rectified and normalized in a consistent way. The white dwarf synthetic spectra and the B-type star spectra used for training and testing of the ANN with their coverage and resolution are as shown in Table 1. The wavelength range of the simulated spectra for the white dwarf stars was 4000-7000 A° with a step of 1 A°, whereas the wavelength range of the spectra for the B-type stars was 3200-10000 A° with a step of 0.01 A°. As a result, it was necessary to compress these datasets to a smaller size by smoothing these data to different wavelength steps and testing them within different neural network configurations. The smoothing process for the data was applied by trying 1001, 601 and 374 data points for the dataset of the simulated spectra which are the inputs to the NN. Figure 4 shows a sample of spectra of these three smoothed configurations for the white dwarf stars, which shows almost typical spectrum flux values.

The smoothing operation suggested here had considerable improvement to the neural network training errors and minimized the training times. In addition to the smoothing process, a normalization process was implemented to the values of wavelength, effective temperature as well as surface gravity such that their values are limited from 0-1 before being fed to the NN for training, verification, or testing. Moreover, the B-type star spectra were trimmed to the wavelength range 4000-7000 A° for the purpose of minimizing the number of data points fed to the NN. This wavelength coverage is more than enough as it contains the important lines necessary to learn the NN to predict the atmospheric parameters.

**Table 1. Synthesized spectra used for training and testing of ANN**.

| Star Class | Number of training spectra | Atmospheric parameters | Range |
|---|---|---|---|
| White Dwarf | 174 | $T_{eff}$ | 50000 – 120000 K, $\Delta T_{eff}$=2500 K |
| | | log g | 7.0 – 8.5, $\Delta$ log g=0.25 |
| B-type Stars | 135 | $T_{eff}$ | 15000- 30000 K, $\Delta T_{eff}$=1000 K |
| | | log g | 1.75 – 4.75, $\Delta$ log g=0.25 |

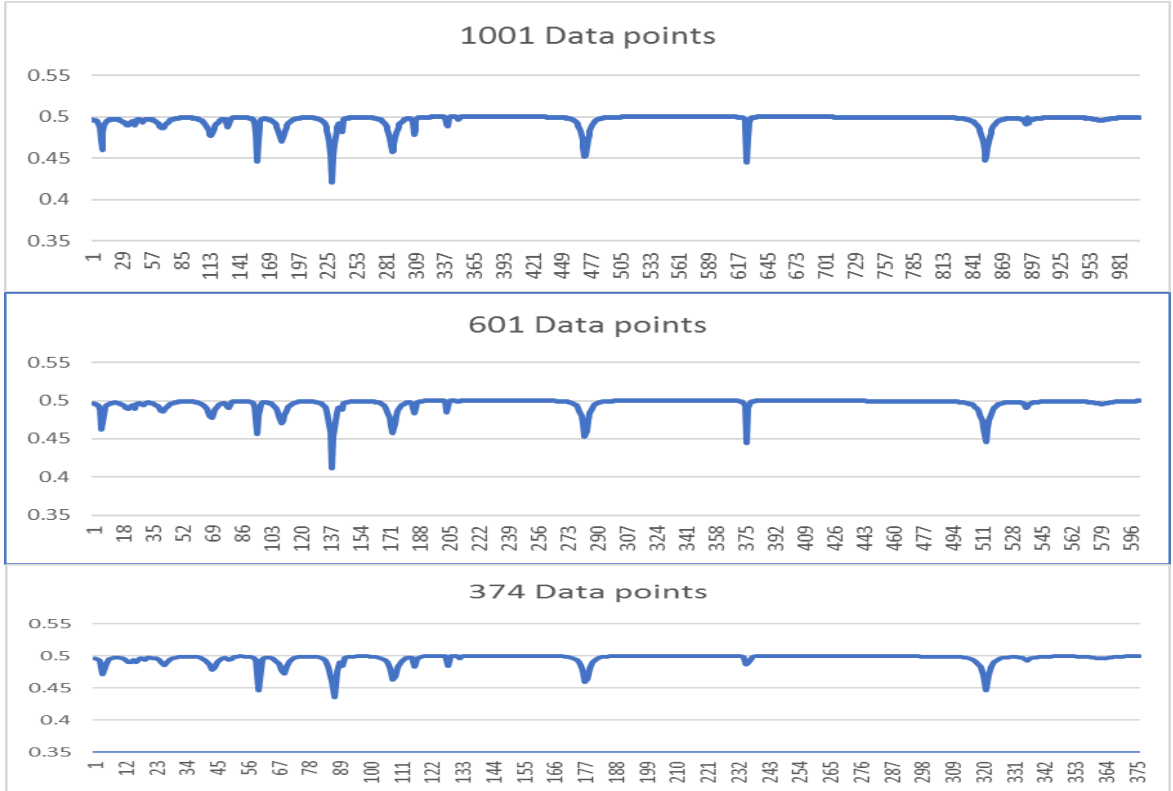

**Figure 4. Smoothing process applied to white dwarf spectra for the purpose of training of the NN**

## 5. Application of ANN for stellar parameterization
### 5.1. White Dwarfs

The training phase of the proposed neural network is implemented by testing three different arrangements for the input values of the network according to the smoothing procedure previously described in section 4.2. In each of these three different arrangements, we tested different neural network configurations for the number of hidden layers in these networks. As a result, we tested NNs with configurations as are shown in table 2. In all of these configurations, we used a single hidden layer with a number of neurons shown in table 2 and used two output nodes in the output layer which are the effective temperature and the surface gravity of the star. The ANN for all of these different configurations has been trained using the backpropagation algorithm (generalized delta rule) previously explained in section 2 with the minimization problem described by equation 6 for the RMS error. The training data we used included 174 DO white dwarf smoothed synthesized spectra that cover the wavelength range 4000-7000 A° with a range of spectral parameters introduced in table 1. Another 29 spectra are left for verification of the NN performance after training. The training stops when the network converges to a minimum value of RMS errors shown in Table 2 and stabilizes there for a long time. The final weights for each configuration are frozen and applied later to verify and test the ANN ability to predict the parameters from unseen spectra. As is shown in this table, the minimum value of RMS error was that of (601-10-2) arrangement which elects it to be the best configuration to be used to predict the atmospheric stellar parameters $T_{eff}$ and $\log g$. During the training process, we used values of learning rate ($\eta$) and momentum (β) shown in table 2.

**Table 2. ANN configurations tested to be used in atmospheric stellar classification for white dwarf**

| ANN configuration | Number of NN input nodes | Number of hidden neurons | Number of output nodes | Learning rate ($\eta$) and Momentum (β) | Training RMS error |
|---|---|---|---|---|---|
| 376-10-2 | 376 | 10 | 2 | 0.2, 0.5 | 0.000238 |
| 376-20-2 | 376 | 20 | 2 | | 0.000243 |
| 376-40-2 | 376 | 40 | 2 | | 0.000252 |
| 601-10-2 | 601 | 10 | 2 | 0.25, 0.5 | 0.000207 |
| 601-20-2 | 601 | 20 | 2 | | 0.000218 |
| 601-40-2 | 601 | 40 | 2 | | 0.005250 |
| 1001-5-2 | 1001 | 5 | 2 | 0.3, 0.5 | 0.000213 |
| 1001-10-2 | 1001 | 10 | 2 | | 0.000541 |
| 1001-20-2 | 1001 | 20 | 2 | | 0.001600 |

Those values for $\eta$ and $\beta$ were found to speed up the convergence of the back-propagation learning algorithm of the ANN without over-shooting the solution. By the end of the training phase, it was necessary to test the effectiveness of the chosen ANN configuration (601-10-2) with respect to other configurations by applying the frozen weights to test and calculate the parameters of the 174 spectra previously used for training. Table 3 shows the RMS error evaluation for the difference between the calculated output parameters of the trained NNs and the desired ones. The error is calculated by the following equations for $T_{eff}$ and $\log g$ receptively.

$$E_{rms}(T_{eff}) = \frac{1}{N}\sqrt{\sum_{N=1}^{N=174}\left(T_{eff\_c} - T_{eff\_d}\right)^2}, \tag{11}$$

$$E_{rms}(log\_g) = \frac{1}{N}\sqrt{\sum_{N=1}^{N=174}\left(log\_g_c - log\_g_d\right)^2}, \tag{12}$$

Where $N$ is the number of training spectra which is 174, $T_{eff\_c}$ is the calculated value of the effective temperature for each spectrum obtained from the trained neural network, and $T_{eff\_d}$ is the desired value of the effective temperature for each spectrum used to train the NN. Similarly, $log\_g_c$ is the calculated value of surface gravity for each spectrum obtained from the trained neural network, and $T_{eff\_d}$ is the desired value of the effective temperature for each spectrum used to train the NN. As is shown in Table 2 the minimum values of the calculated errors were for the NN which has the (601-10-2) arrangement. This assures that this arrangement is the best configuration for the neural network that we will use to predict the atmospheric parameters of the unknown white dwarf spectrum.

**Table 3. RMS error of the calculated values previously used in training of the ANN**

| ANN configuration | $E_{rms}(T_{eff})$ | $E_{rms}(log\_g)$ |
|---|---|---|
| 376-10-2 | 69.9 | 0.007 |
| 376-20-2 | 68.9 | 0.009 |
| 376-40-2 | 78.1 | 0.009 |
| 601-10-2 | 47 | 0.005 |
| 601-20-2 | 55.6 | 0.011 |
| 601-40-2 | 2461 | 0.053 |
| 1001-5-2 | 58.6 | 0.01 |
| 1001-10-2 | 117.8 | 0.006 |
| 1001-20-2 | 310 | 0.005 |

To verify the trained ANN algorithm, we computed the fundamental parameters for 29 synthetic models not previously used in the training of the network by the use of the trained network with the (601-10-2) configuration which gives the effective temperatures and surface gravity pairs listed in Table 4. As shown in Table 4, there is very good agreement between the input and predicted models.

**Table 4. Verification of ANN use to predict the 29 white dwarfs synthetic models**

| Log_g_601_10_2 | Log_g | Teff_601_10_2 | Teff |
|---:|---:|---:|---:|
| 8.471779 | 8.5 | 50177.31 | 50000 |
| 8.481155 | 8.5 | 52604.05 | 52500 |
| 8.471093 | 8.5 | 55176.85 | 55000 |
| 8.466984 | 8.5 | 57669.56 | 57500 |
| 8.465737 | 8.5 | 60105.53 | 60000 |
| 8.465545 | 8.5 | 62510.49 | 62500 |
| 8.465741 | 8.5 | 64886.47 | 65000 |
| 8.466002 | 8.5 | 67279.61 | 67500 |
| 8.466006 | 8.5 | 69663.76 | 70000 |
| 8.465481 | 8.5 | 72060.55 | 72500 |
| 8.464399 | 8.5 | 74541 | 75000 |
| 8.462972 | 8.5 | 77156.85 | 77500 |
| 8.460953 | 8.5 | 79878.66 | 80000 |
| 8.458674 | 8.5 | 82700.87 | 82500 |
| 8.456223 | 8.5 | 85528.73 | 85000 |
| 8.453735 | 8.5 | 88273.54 | 87500 |
| 8.451658 | 8.5 | 90900.53 | 90000 |
| 8.450181 | 8.5 | 93391.86 | 92500 |
| 8.449382 | 8.5 | 95738.95 | 95000 |
| 8.449507 | 8.5 | 97994.3 | 97500 |
| 8.450131 | 8.5 | 100501.8 | 100000 |
| 8.451875 | 8.5 | 102757.1 | 102500 |
| 8.454079 | 8.5 | 105097.5 | 105000 |
| 8.456409 | 8.5 | 107572.2 | 107500 |
| 8.458573 | 8.5 | 110156.1 | 110000 |
| 8.460049 | 8.5 | 112779.1 | 112500 |
| 8.460311 | 8.5 | 115248.1 | 115000 |
| 8.458571 | 8.5 | 117300 | 117500 |
| 8.453732 | 8.5 | 118703.9 | 120000 |

Now we turn to apply the code on the observed spectra of some helium rich white dwarfs. We used the observed spectra of the DO white dwarfs from the Data Release Four (DR4) of SDSS. The SDSS is a photometric and spectroscopic survey covering 700 square degrees of the sky around the northern Galactic cap (Adelman-McCarthy et al., 2005). The main goal of the survey was to study the large-scale structure of the universe. A small fraction of the observed stars is targeted for spectroscopy. The resulting data are in low resolution (R = 1800, FWHM $\simeq$ 3 A$^0$). The flux calibrated spectra cover the range between 3800 and 9000 A$^o$.

We selected 13 candidates and compared the derived fundamental parameters with that deduced by the MDM method. We list the results in Table 5, where column 2 represents the effective temperatures predicted by the present ANN algorithm, column 3 is the effective temperature predicted from the MDM method, column 4 is the surface gravity predicted from the ANN algorithm and column 5 is the surface gravity predicted from the MDM method.

In Figures 5 and 6, we plotted the effective temperatures and surface gravities predicted from the ANN versus those predicted from the MDM. In general, there is a good agreement between the results for seven stars, and intermediate discrepancies for the rest six stars. As shown, the discrepancies are larger in the surface gravities than that of the effective temperatures. These discrepancies may be attributed to that, there are few lines in the spectrum of the DO white dwarfs that make the training process of the ANN difficult and less accurate.

**Table 5: Fundamental Parameters of the DO white dwarfs**.

| Star name | $T_{eff}$ (ANN) K | $T_{eff}$ (MDM) K | log g (ANN) | log g (MDM) |
|---|---|---|---|---|
| SDSS J034101.39+005353.0 | 59948.89 | 60000 | 7.72 | 7.750 |
| SDSS J034227.62+072213.2 | 53156.07 | 52500 | 7.50 | 7.750 |
| SDSS J075540.94+400918.0 | 103122.30 | 102500 | 7.047 | 7.250 |
| SDSS J084008.72+325114.6 | 97610.02 | 87500 | 7.51 | 8.000 |
| SDSS J091433.61+581238.1 | 118066.50 | 110000 | 7.09 | 8.000 |
| SDSS J102327.41+535258.7 | 119978.19 | 120000 | 8.01 | 8.000 |
| SDSS J113609.59+484318.9 | 49440.19 | 50000 | 7.97 | 8.000 |
| SDSS J134341.88+670154.5 | 90148.43 | 100000 | 8.09 | 7.750 |
| SDSS J144734.12+572053.1 | 118564.87 | 102500 | 8.3 | 7.750 |
| SDSS J154752.33+423210.9 | 83975.30 | 95000 | 7.54 | 7.250 |
| SDSS J155356.81+433228.6 | 75769.56 | 77500 | 7.80 | 7.500 |
| SDSS J204158.98+000325.4 | 119634.49 | 100000 | 7.58 | 7.250 |
| SDSS J140409.96+045739.9 | 72629.26 | 70000 | 7.5 | 7.500 |

### 5.2 B-type Stars

As previously described for the white dwarfs in section 5.1, the same techniques are used to decide about the best configuration of the ANN that can be used to predict the atmospheric parameters of B-type stars from their stellar spectra. Table 6 shows the configurations for the ANNs that have been tested for the sake of the best classification network. Similar smoothing criteria have been implemented for the 135 synthesized spectra used to train different ANNs shown in Table 6. The same backpropagation algorithm (generalized delta rule) has been used to train different neural networks and the training stopped when the network converged to a minimum value of RMS error shown in Table 6 and stabilize there for a long time. As is shown in this table, the minimum value of RMS error was that of the (601-10-2) arrangement (which was expected) to be the best configuration for the prediction of the atmospheric stellar parameters $T_{eff}$ and $log\_g$ of B-type stars.

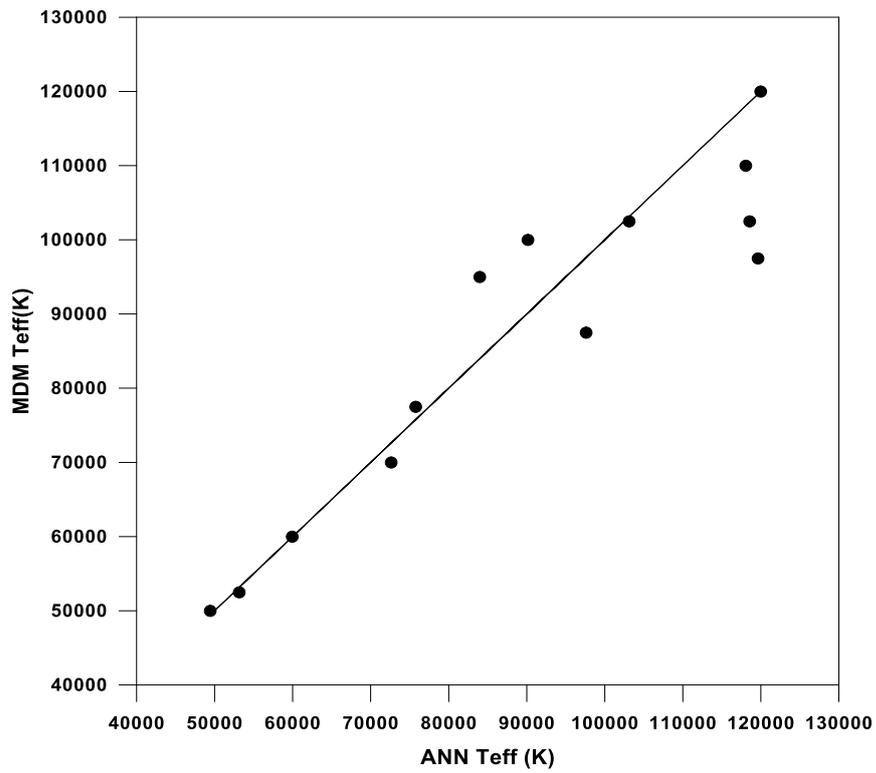

**Figure 5. Comparison between the predicted effective temperatures computed by ANN and MDM algorithms.**

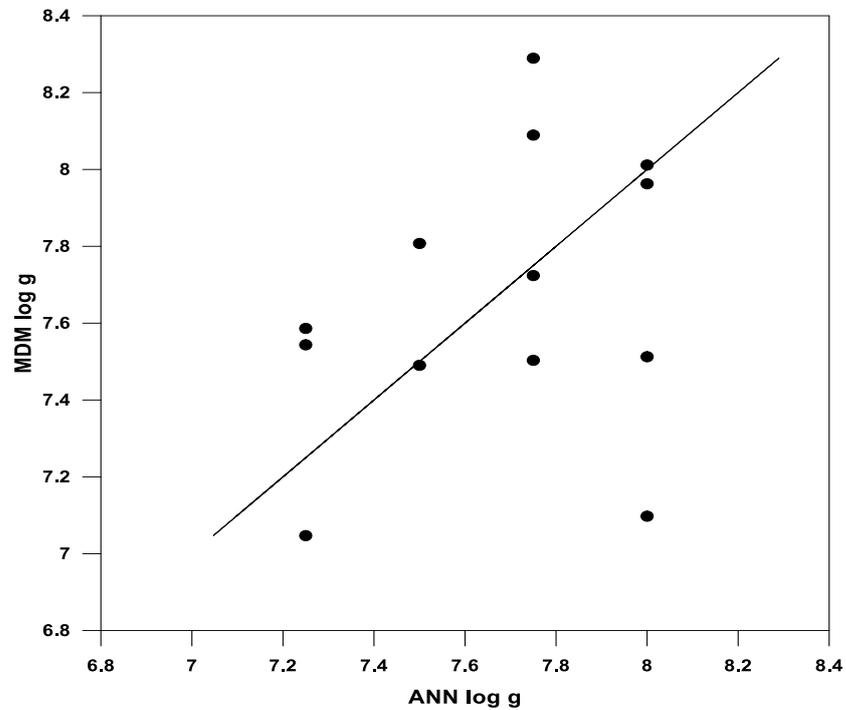

**Figure 6. Comparison between the predicted surface gravity computed by ANN and MDM algorithms**.

Same techniques described in section 5.1 for the test of the effectiveness of the chosen ANN configuration (601-10-2) with respect to other configurations by applying the final frozen weights of the training phase to calculate the parameters of the 135 spectra used for training. Table 7 shows the RMS error evaluation for the difference between the calculated output parameters of the trained NNs and the desired ones which are calculated by Equations (11) and (12) for $T_{eff}$ and $log\ g$ respectively. Again, the best configuration of the ANN is that of the (601-10-2) arrangement.

**Table 6. ANN configurations tested to be used in atmospheric stellar classification for B-type stars**

| ANN configuration | Number of NN input nodes | Number of hidden neurons | Number of output nodes | Learning rate ($\eta$) and Momentum ($\beta$) | Training RMS error |
|---|---|---|---|---|---|
| 376-10-2 | 376 | 10 | 2 | 0.25, 0.5 | 0.000269 |
| 376-20-2 | 376 | 20 | 2 | | 0.000124 |
| 376-40-2 | 376 | 40 | 2 | | 0.001220 |
| 601-10-2 | 601 | 10 | 2 | 0.3, 0.5 | 0.00006 |
| 601-20-2 | 601 | 20 | 2 | | 0.000163 |
| 601-40-2 | 601 | 40 | 2 | | 0.007000 |
| 1001-5-2 | 1001 | 5 | 2 | 0.2, 0.5 | 0.00300 |
| 1001-10-2 | 1001 | 10 | 2 | | 0.007065 |

**Table 7. RMS error of the calculated values previously used to train ANN for B-Type stars**

| ANN configuration | $E_{rms}\ (T_{eff})$ | $E_{rms}\ (Log\_g)$ |
|---|---|---|
| 376-10-2 | 31.87 | 0.474 |
| 376-20-2 | 21.52 | 0.097 |
| 376-40-2 | 777 | 0.91 |
| 601-10-2 | 4.72 | 0.081 |
| 601-20-2 | 36.54 | 0.206 |
| 601-40-2 | 4138 | 29.2 |
| 1001-5-2 | 568 | 1.95 |
| 1001-10-2 | 768.3 | 6.5 |

As we did for the white dwarf spectra, we verified the effectiveness of ANN trained algorithm by computing the fundamental parameters for 22 B-type synthetic spectra not previously used in the training process which give the effective temperatures and surface gravity pairs listed in Table 8. As is shown in Table 8, there is a very good agreement between the input and predicted models.

**Table 8. Verification of ANN use to predict the 22 B-Type synthetic models**

| Log_g_600_10_2 | Log_g | Temp_600_10_2 | Teff |
|---|---|---|---|
| 3.744552 | 3.75 | 16042.66 | 16000 |
| 2.991446 | 3.00 | 17986.92 | 18000 |
| 2.761616 | 2.75 | 19989.89 | 20000 |
| 3.006225 | 3.00 | 20946.19 | 21000 |
| 3.505322 | 3.50 | 21958.18 | 22000 |
| 3.741411 | 3.75 | 22943.37 | 23000 |
| 4.240611 | 4.25 | 24993.91 | 25000 |
| 4.259135 | 4.25 | 28006.74 | 28000 |
| 4.514815 | 4.5 | 28042.88 | 28000 |
| 4.740483 | 4.75 | 28092.16 | 28000 |
| 3.012029 | 3 | 29171.13 | 29000 |
| 3.265319 | 3.25 | 29126.43 | 29000 |
| 3.514966 | 3.5 | 29128.49 | 29000 |
| 3.759563 | 3.75 | 29071.65 | 29000 |
| 4.005296 | 4 | 29019.89 | 29000 |
| 4.259472 | 4.25 | 29016.79 | 29000 |
| 4.51338 | 4.5 | 29029.15 | 29000 |
| 4.738731 | 4.75 | 29061.75 | 29000 |
| 3.016044 | 3 | 30072.63 | 30000 |
| 3.260371 | 3.25 | 30025.06 | 30000 |
| 3.508059 | 3.5 | 30057.52 | 30000 |
| 3.748828 | 3.75 | 30008.37 | 30000 |

We used the observed spectra for B-type stars retrieved from the archive of 2-m Telescope's Cassegrain lens at the Ondrejov observatory. The spectra under investigation were mostly taken from the HEROS spectra which were obtained using Echelle spectrograph HEROS, Kubat et al. (2010), Saad and Nouh (2011), Saad and Nouh (2011), Nouh et al. (2013). The spectrum spans the range of wavelengths 3450 A ° -8620A °, the Balmer lines up to H15, as well as some infrared lines. The resolution is R= 20000, which is equivalent to FWHM= 0.25 A °. Table 9 shows the list of observed objects and their predicted parameters, and Figures 7 and 8 plot the comparison between the effective temperatures and gravities obtained by the ANN and MDM methods. The predicted parameters from ANN and MDM are in good agreement except the surface gravities of the star ρ Aur.

**Table 9: Fundamental parameters of the B-type stars.**

| Star | Teff (MDM) | Log g (MDM) | Teff (ANN) | log g (ANN) |
|---|---|---|---|---|
| 96 her | 17000 | 4 | 17038 | 4.002 |
| α Vir | 25000 | 3.75 | 24461.6 | 3.612 |
| β CMa | 24000 | 3.75 | 22282.6 | 3.567 |
| $\varepsilon$ Per | 30000 | 4 | 25201.7 | 3.512 |
| i her | 17000 | 3.75 | 19838 | 3.939 |
| ρ Aur | 15000 | 4 | 13943.2 | 2.293 |
| U her | 19000 | 3.5 | 18205.9 | 3.621 |

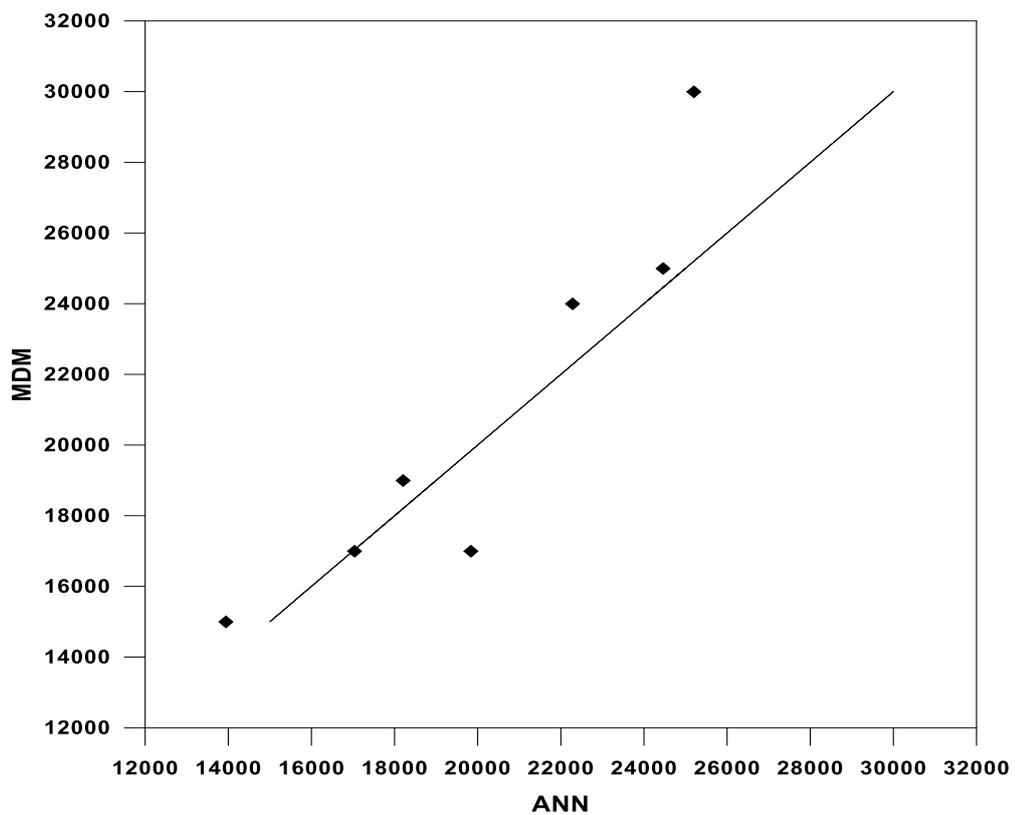

**Figure 7. Comparison between the predicted effective temperatures computed by ANN and MDM algorithms for B-type stars.**

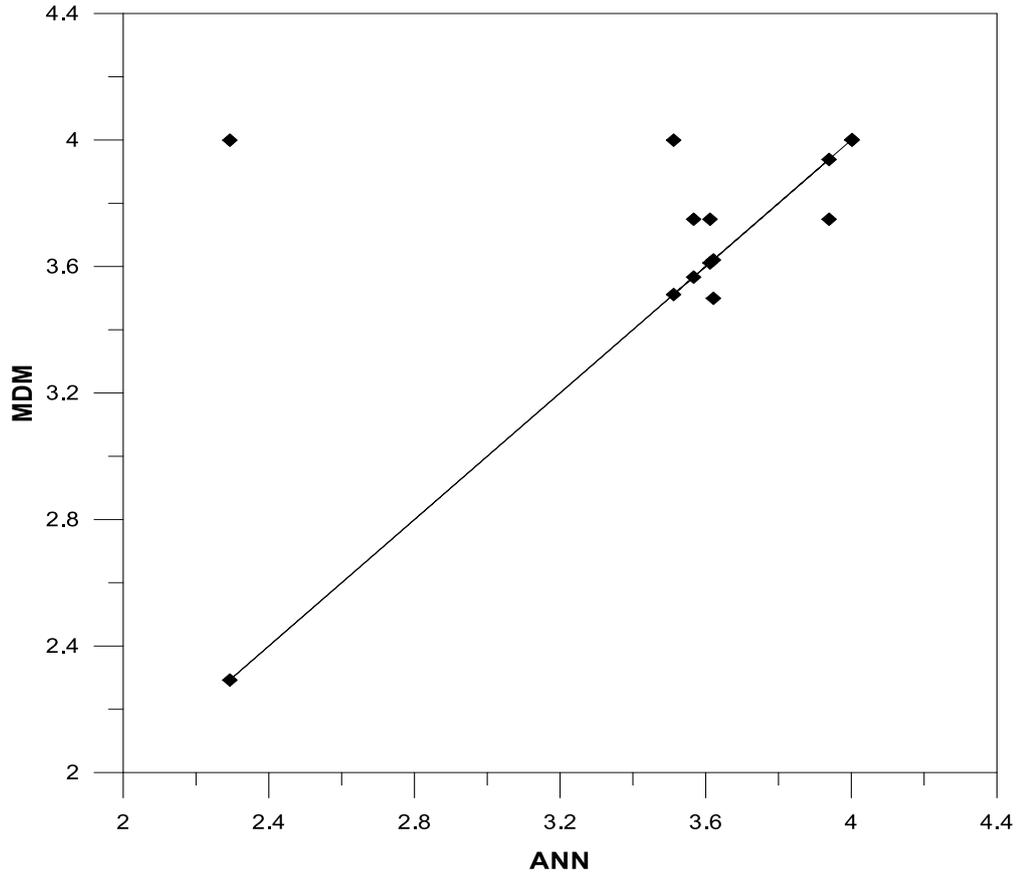

**Figure 8. Comparison between the predicted surface gravity computed by ANN and MDM algorithms for B-type stars**.

## 6. Conclusion

We developed an Artificial Neural Network (ANN) algorithm for automated classification of stellar spectra. For the purpose of testing the algorithm, we derived the fundamental parameters for a total of 29 theoretical spectra and 13 observed spectra of the white dwarf stars and a total of 22 theoretical spectra and 7 observed spectra of the B-type stars. The comparison between the ANN and MDM is satisfactory for most of the tested spectra. In summarizing the results, we inferred that artificial neural networks are an outstanding computational choice for very large projects. In smaller projects where better control is necessary, minimum distance approaches that take advantage of interpolation and optimization may be more efficient and versatile. Also, ANN is somewhat more rigid than MDM methods. If a neural network is trained to search for all parameters, a variation of the problem that may use additional information to constrain, and the only search for the remaining parameters will take a new ANN to be trained. Another problem encountered when using ANN is the missing data found in the observed spectra, the problem which not appeared in the minimum distance method.